\documentclass[journal=jpclcd,manuscript=letter,layout=traditional]{achemso}

\usepackage{amssymb}
\usepackage{amsmath}

\usepackage{graphicx} 
\usepackage{wrapfig}

\usepackage{xcolor}
\usepackage{multirow}

\usepackage{xr}

\newcommand*{\rom}[1]{\expandafter\@slowromancap\romannumeral #1@}

\author{Nikhil V. S. Avula}
\email{nikhil@jncasr.ac.in}
\affiliation{Chemistry and Physics of Materials Unit, Jawaharlal Nehru Centre for Advanced Scientific Research, Bangalore 560064, India}

\author{Michael L. Klein}
\affiliation{Institute for Computational Molecular Science, Temple University, Philadelphia, United States}

\author{Sundaram Balasubramanian}
\email{bala@jncasr.ac.in}
\phone{+91 (80) 2208 2808}
\fax{+91 (80) 2208 2766}
\affiliation{Chemistry and Physics of Materials Unit, Jawaharlal Nehru Centre for Advanced Scientific Research, Bangalore 560064, India}

\title[ML-electrolyte]{Understanding the Anomalous Diffusion of Water in Aqueous Electrolytes Using Machine Learned Potentials}

\begin{document}

\begin{tocentry}
    \includegraphics[width=3.25in]{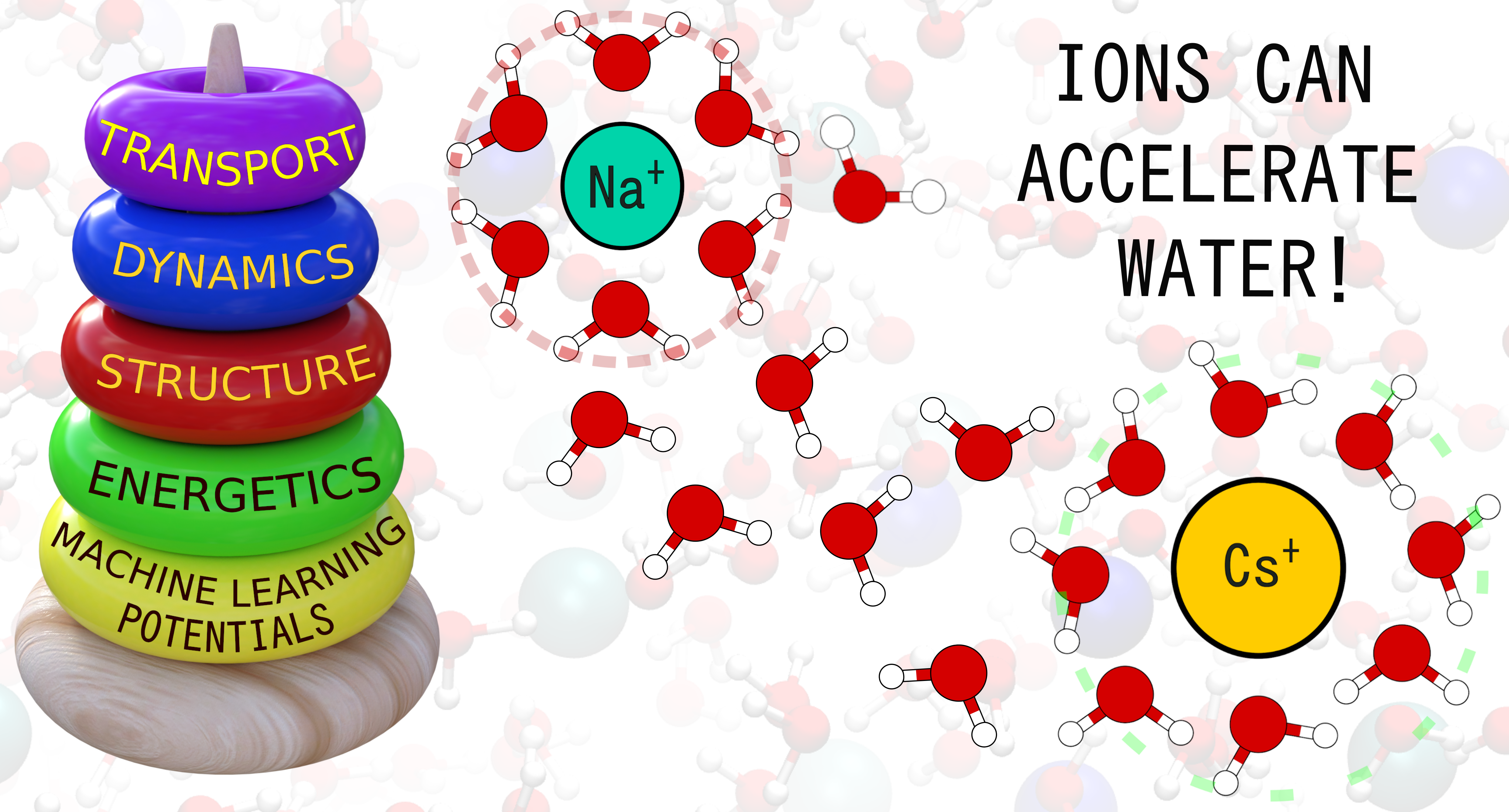}
    For Table of Contents Only.
\end{tocentry}

\begin{abstract}
The diffusivity of water in aqueous cesium iodide solutions is larger than that
in neat liquid water, and vice versa for sodium chloride solutions. Such
peculiar ion-specific behavior, called anomalous diffusion, is not
reproduced in typical force field-based molecular dynamics (MD) simulations due to inadequate
treatment of ion-water interactions. Herein, this hurdle is tackled using
machine learned atomic potentials (MLPs) trained on data from density functional
theory calculations. MLP-based atomistic MD simulations of aqueous salt
solutions reproduce experimentally determined thermodynamic, structural,
dynamical, and transport properties, including their varied trends of water
diffusivities across salt concentration. This enables an examination of their
intermolecular structure to unravel the microscopic underpinnings of the
distinction in their transport. While both ions in CsI solutions contribute to
faster diffusion of water molecules, the competition between the heavy
retardation by Na-ions and slight acceleration by Cl-ions in NaCl solutions
reduces their water diffusivity. 
\end{abstract}

Aqueous electrolyte solutions are widely prevalent and seemingly simple, yet
display diverse transport properties
\cite{bagchi_PRL_1995,chandra_PRL_2000,kdill_JACS_2002,collins_methods_2004,fayer_pnas_2007,marcus_chemrev_2009,laage_JACS_2013,panagiotopolous_JCP_2020,bagchi_JCP_2020,mundy_AccChemRes_2021,panagiotopolous_JPCB_2023}.
An interesting phenomenon they exhibit is the ``anomalous diffusion'' of water,
which is particularly difficult to capture in molecular simulations
\cite{yethiraj_JPCB_2012,parrinello_PNAS_2014}. Although salts dissolved in
water are expected to make the solution more viscous (kosmotropes), there exist
many that make the solutions less viscous (chaotropes). While the general
behavior of salt solutions can be rationalized using the Hofmeister series of
their respective ions, a microscopic understanding of transport is still
evolving\cite{collins_methods_2004,marcus_chemrev_2009,kdill_chemrev_2017,hofmeister_thompson_JPCB_2021}. In this context, accurate molecular simulations which can
quantitatively reproduce experimental data can yield vital insights to unravel
the structure-dynamics-property paradigm.

While aqueous salt solutions modeled with classical force fields (either with
or without explicit atom/ion polarizabilities) reproduce the behavior of
kosmotropes such as NaCl, they fail to do so for chaotropes such as CsI
\cite{yethiraj_JPCB_2012,panagiotopolous_MolPhys_2019}. Parrinello and
co-workers used density functional theory (DFT) based ab initio molecular
dynamics (AIMD) simulations of 3 M NaCl and CsI solutions to study this
phenomenon and concluded that the lack of dynamical heterogeneity in
non-polarizable force field (NPFF) simulations is the reason for their failure
\cite{parrinello_PNAS_2014}. Berkowitz and co-workers concluded that a
fluctuating charge force field with an additional dynamical charge transfer
(FQ-DCT) term is essential to reproduce the experimentally observed
acceleration of water in aqueous KCl (mildly chaotropic) solutions
\cite{berkowitz_JPCL_2014,berkowitz_JCP_2015}. In a recent perspective,
Panagiotopoulos and Yue delineated the challenges faced by molecular
simulations in the quantitative description of the dynamics in aqueous
electrolyte solutions \cite{yue_thesis_2021,panagiotopolous_JPCB_2023}. They
also illustrated the prospects for machine learned potentials (MLPs) trained on ab
initio datasets to qualitatively capture the different behavior of water
diffusion in four alkali halide solutions. However, a clear microscopic
explanation for the phenomenon is yet to be forthcoming. This is the task we
set out to achieve. 

In this letter, we demonstrate that Machine Learned potentials trained
on DFT data can capture the anomalous diffusion phenomenon and overcome the
challenges faced by both classical (poor accuracy) and ab initio MD
(short trajectories) simulations. Furthermore, we relate the trends in
the transport properties to differences in the solvation shell structure and
dynamics, thereby establishing a structure-property relationship. To this end, we
study aqueous NaCl and CsI solutions with specific focus on the behavior of
water diffusivity across salt concentration. First, we benchmark the MLPs on
various structural, thermodynamic and transport properties of neat water and
salt solutions. This is an important step to not only validate the MLPs but
also the underlying DFT functional used to train them. Subsequently, we show
that the trends in diffusivity and shear viscosity of salt solutions from MLPs
match that of experiments qualitatively and are hence able to distinguish
between the NaCl and CsI solutions. Finally, we show that NPFFs do not capture
the qualitative differences in the solvation shells of Na and Cs ions, and
hence fail to reproduce their varied transport behavior. 

\begin{table}
\centering
\caption{\label{tab:benchmark} Comparison of thermodynamic, structural,
	dynamical and transport properties from DPMD against the corresponding
	experimental values. Unless specified otherwise, all the properties
	reported are at 300 K and 1 bar. Wherever available, the uncertainty in
	the last digit of the estimates are mentioned in parenthesis. $^{*}$cn:
	coordination number.}

\begin{tabular}{cccc}
\hline
\hline
Property                                 & DPMD  & Experiment                                 & Error (\%) \\ 
\hline
Neat water density (g/cc)                & 0.972 & 0.997\cite{water_density_nist_2001}        & -2.5 \\
O-O g(r) - first max r ($\mathrm{\AA}$)  & 2.80  & 2.80(1)\cite{oogofr_skinner_JCP_2013}      &  -  \\
O-O g(r) - first max g(r)                & 2.64  & 2.57(5)\cite{oogofr_skinner_JCP_2013}      & 2.7  \\
O-O g(r) - first min r ($\mathrm{\AA}$)  & 3.48  & 3.45(4)\cite{oogofr_skinner_JCP_2013}      & 0.9 \\
O-O g(r) - first min g(r)                & 0.83  & 0.84(2)\cite{oogofr_skinner_JCP_2013}      & -1.2 \\
O-O cn$^{*}$ at 3.3 $\mathrm{\AA}$       & 4.39  & 4.3(1)\cite{oogofr_skinner_JCP_2013}       & 2.1 \\
Neat water D (10$^{-5}$ cm$^2$/s)        & 2.31  & 2.299\cite{hertz_exp_diffusion_JPC_1996}   & 0.5 \\
Neat water shear viscosity (cP)          & 0.90  & 0.89 \cite{kestin_naclvisc_1977}           & 1.1 \\
\hline
\hline
\end{tabular}
\end{table}

\begin{figure}[h!]
    \includegraphics[width=0.35\linewidth]{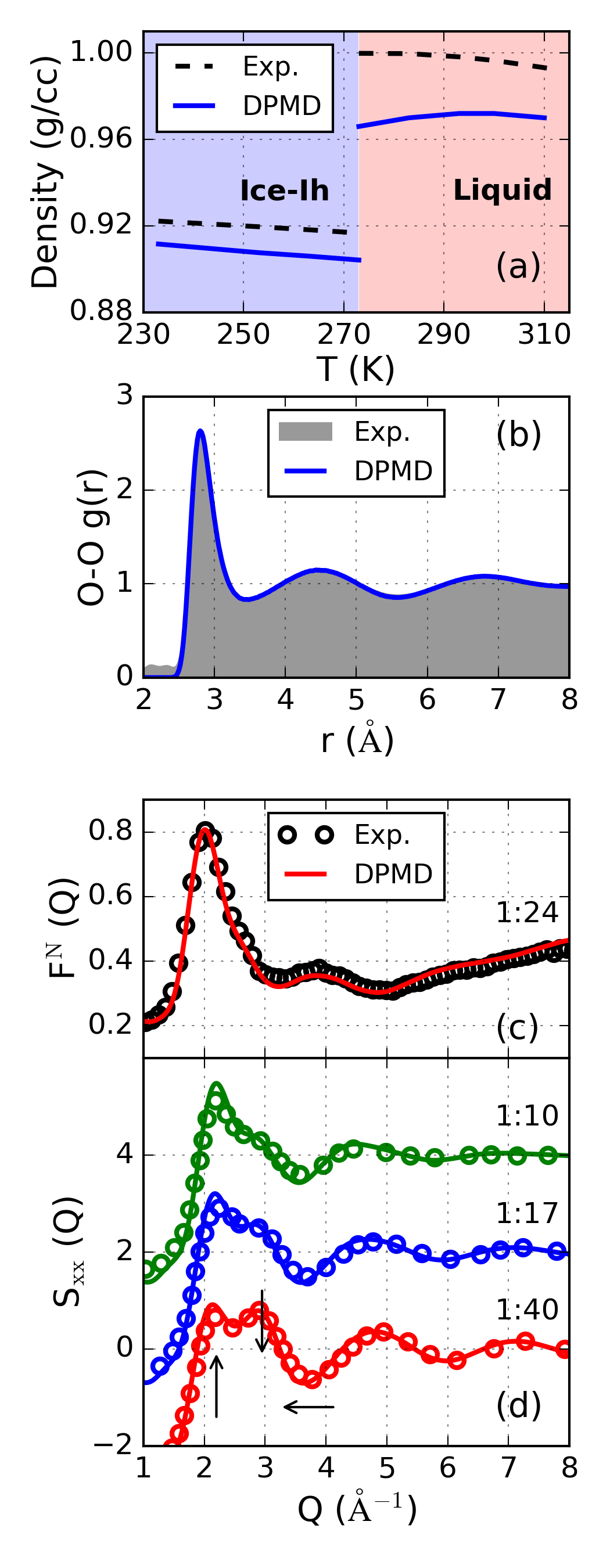}
    \caption{(a) Temperature dependence of density of neat liquid water and
	ice-\textrm{I}h from DPMD and experiments
	\cite{water_density_nist_2001,ice_density_wagner_JPCRD_2006,ice_density_nist_2019}.
	(b) O-O radial distribution function in liquid water at 300 K from DPMD
	and experiments \cite{oogofr_skinner_JCP_2013}. (c) neutron structure
	factor of 2.3 m aqueous CsI solution at 300 K
	\cite{csi_neutron_JPCB_2012}. The concentrations of the salt solutions
	are indicated as salt to water molar ratios. (d) partial neutron
	structure factor S$_{\mathrm{xx}}$(Q) of NaCl solutions at 300 K
	\cite{soper_nacl_neutron_JPCB_2007}. S$_{\mathrm{xx}}$(Q) is defined as
	$\sum_{\alpha,\beta \neq \mathrm{H}} w_{\alpha\beta} S_{\alpha\beta}
	\mathrm{(Q)}$, where $w_{\alpha\beta}$ is the composition-dependent
	weight and  $S_{\alpha\beta} \mathrm{(Q)}$ the partial structure factor
	corresponding to the species $\alpha,\beta$. The structure factor plots
	of 1:17 and 1:10 NaCl solutions are shifted up by two and four units
	respectively, for visual clarity. Arrows indicate the changes induced
	by the addition of NaCl, originally pointed out by Soper et al.
	\cite{soper_nacl_neutron_JPCB_2007} (See SI section
	S3.2 for additional details.)}
	\label{fig:benchmark}
\end{figure}

To this end, two MLP force fields, one each for NaCl and CsI
systems were constructed using the Deep Potential (DP) framework
\cite{deepmdkit_CPC_2018,deepmdse_NIPS_2018,deepmdkit_arxiv_2023} trained on
condensed phase DFT data using the revPBE-D3 functional
\cite{PBE_1996,revPBE_1998}. A comparison of several thermodynamic, structural,
and transport properties computed from DP based MD simulations
(DPMD) with corresponding experimental data demonstrates its efficacy (Figure
\ref{fig:benchmark}, Table \ref{tab:benchmark}, and Supporting Information (SI)
section S2). Figure \ref{fig:benchmark} (a) shows the
temperature dependence of the density of liquid water and ice. DPMD captures
the major qualitative trends including - (i) liquid water’s non-monotonic
dependence and (ii) the large difference in the ice-water density at 273 K
(figure \ref{fig:benchmark}(a)). This result is particularly impressive because
the ice snapshots were not included in the training of the MLPs and hence
demonstrates their robustness. Figure \ref{fig:benchmark}(b) compares the
experimental and DPMD water structure using the O-O radial distribution
function (g(r)) and shows a good agreement between them including the major
peak positions and heights (Tables \ref{tab:benchmark} and S3).  

Figures \ref{fig:benchmark} (c) and (d) compare the neutron structure factors
of aqueous CsI and NaCl solutions obtained from experiments and our DPMD
simulations, again showing a good correspondence. The arrows in figure
\ref{fig:benchmark} (d) indicate the qualitative changes that the addition of
NaCl salt induces to the water structure and are similar to the trends reported
by Soper et al from their experiments \cite{soper_nacl_neutron_JPCB_2007}.
Using SCAN based DPMD simulations, Zhang et al recently showed that the changes
in the water structure due to the addition of salts are qualitatively different
from that brought about by external pressure
\cite{zhang_salt_pressure_natcomm_2022}. Our results are consistent with their
observations. Table \ref{tab:benchmark} displays the quantitative accuracy of
DPMD in estimating transport properties such as the self-diffusion coefficient
(D), and shear viscosity ($\eta$) of liquid water. Further, the hydration shell
properties of the ions were also compared with experimental and ab initio
simulation results (see SI section S2.3). Having
extensively benchmarked both structural and transport properties estimated by
DPMD simulations, we now use them to investigate the “anomalous diffusion” of
water in aqueous salt solutions. 

\begin{figure}[h!]
    \includegraphics[width=1.0\linewidth]{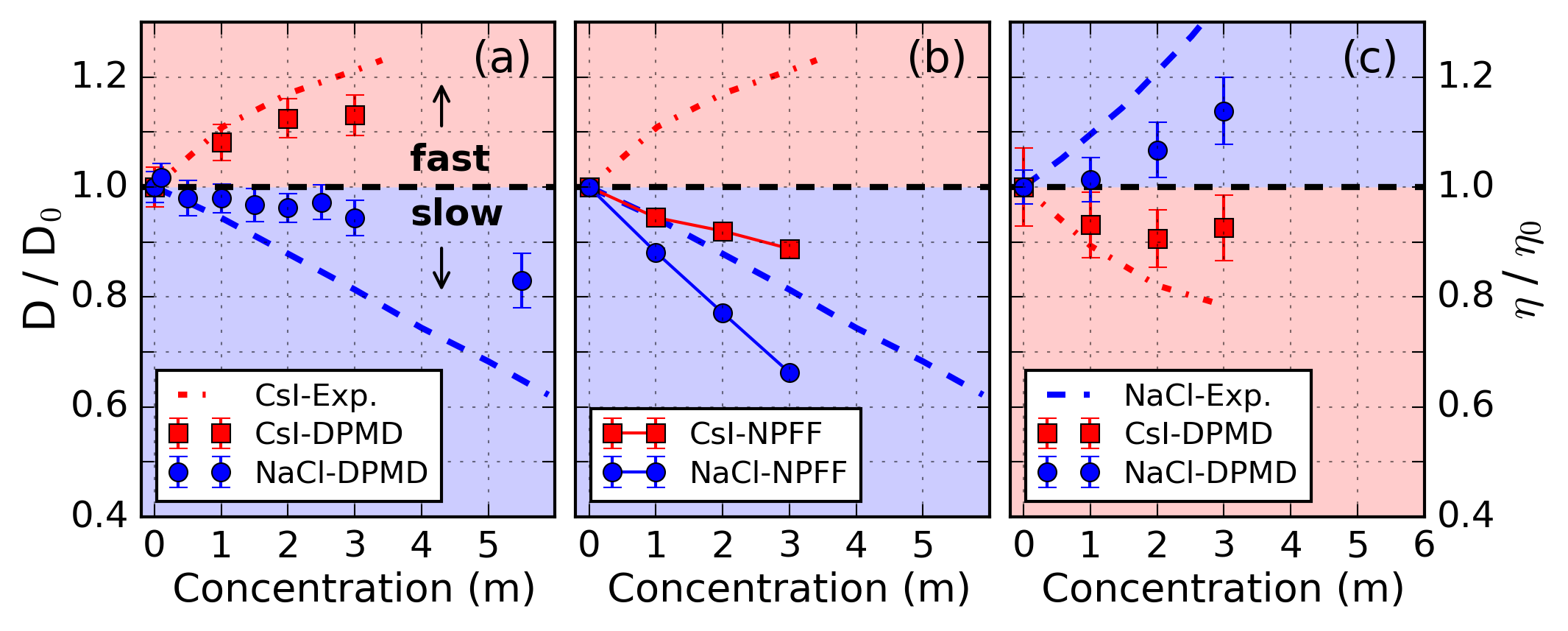}
    \caption{(a) Ratio of self-diffusion coefficient of water molecules in salt
	solutions to those in neat water at 300 K from DPMD and experiments.
	(b) self-diffusivity ratio obtained from NPFF simulations at 300 K. (c)
	shear viscosity ratio of salt solutions estimated using DPMD
	simulations at 300 K. Experimental values of self-diffusion
	coefficients are taken from Ref \cite{hertz_exp_diffusion_JPC_1996} and
	those of shear viscosity are taken from Refs \cite{jones_JACS_1936},
	\cite{kestin_naclvisc_1977} . In all the plots, the red region
	indicates faster dynamics and the blue region indicates slower
	dynamics. The standard errors are equal to 2$\sigma$ (twice the
	standard deviation of the bootstrap population) giving a confidence
	interval of about 95\%.} \label{fig:diff-ratio}
\end{figure}

Figure \ref{fig:diff-ratio} (a) shows the relative self-diffusion coefficients
(D/D$_0$) of water molecules in CsI and NaCl solutions obtained from DPMD
simulations. D/D$_0$ values in CsI solutions are greater than unity, indicating
the faster (when compared to neat liquid water) translational diffusion of
water molecules in these solutions even at concentrations as high as 3 m.  This
significant increase in D/D$_0$ in chaotropic salt solutions like CsI, termed
as “anomalous diffusion”, has so far eluded most force field based molecular
simulations. For instance, D/D$_0$ values estimated from the non-polarizable
Madrid-2019 force field based classical MD simulations are below unity at all
concentrations, indicating a slowdown contrary to experimental observations
(figure \ref{fig:diff-ratio} (b))
\cite{vega_madrid2019_JCP_2019,vega_madrid2019_JCP_2022}. This deficiency is not
specific to the Madrid-2019 force field but to almost all classical non-polarizable
force fields (and even to some polarizable force fields)
\cite{yethiraj_JPCB_2012}.  

On the other hand, the D/D$_0$ values in NaCl solutions estimated from DPMD
simulations are lesser than unity, indicating a slowdown, qualitatively
consistent with experiments (see Figure \ref{fig:diff-ratio} (a)). Classical
force fields too correctly capture the slowdown qualitatively, although the
extent of slowdown is exaggerated as seen in Figure \ref{fig:diff-ratio} (b).
While the self-diffusion coefficient (D) is a measure of individual molecular
motion in the fluid, shear viscosity ($\eta$) is a collective property that
depends on all the components of the fluid (including the ions). Figure
\ref{fig:diff-ratio} (c) shows the relative viscosity ($\eta/\eta_0$) estimated
from DPMD simulations with values lesser and greater than unity for CsI and
NaCl solutions respectively, consistent with experiments. In contrast,
$\eta/\eta_0$ values estimated from classical NPFF simulations are less than
unity (slowdown) for both the salt solutions pointing to their inadequacy
\cite{vega_madrid2019_JCP_2019,vega_madrid2019_JCP_2022}. 

To understand the microscopic origins of the anomalous diffusion behavior, we
first examine the differences in the hydration shell structure of Cs and Na
ions using radial distribution functions (see SI section S4.1). 
The ion-O g(r)s (Figures S13, S14) in the NaCl solution are more
structured than in the CsI solutions, with a taller first peak and a deeper
first minimum, and in good agreement with previous literature reports
\cite{soper_nacl_neutron_JPCB_2007,csi_neutron_JPCB_2012,cs_paesani_JPCL_2019,na_k_paesani_JPCB_2022,wang_hydshell_JPCL_2023}.
We further investigate the radial structure of the hydration shells of the ions
using incremental g(r)s (iRDFs) which split the total g(r) into contributions
from the N-th oxygen atom sorted by their distance to the ions (see SI section
S3.3 for the definition). iRDFs (and related incremental
probability density functions iPDFs) have been employed earlier to study the
hydration structure in aqueous electrolyte solutions
\cite{mundy_AccChemRes_2021,iodine_mundy_JPCB_2010,iodate_mundy_JPCL_2011,cs_paesani_JPCL_2019,na_k_paesani_JPCB_2022}.  

\begin{figure}[h!]
    \includegraphics[width=1.0\linewidth]{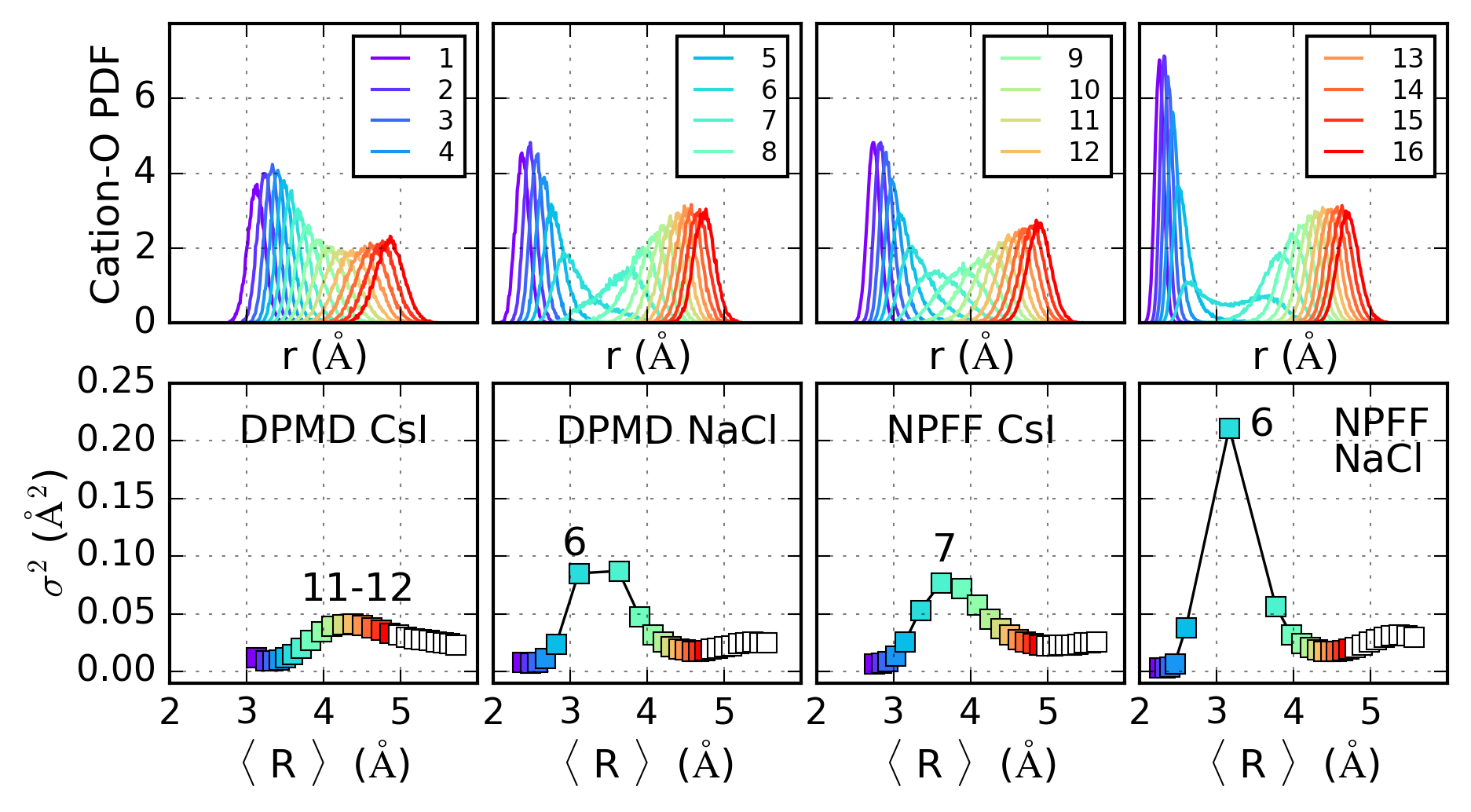}
    \caption{Top Row: Incremental probability density functions (defined in SI
	section S3.3) of the first sixteen nearest neighbor O
	atoms from cations (Na/Cs) in 0.1 m aqueous solutions at 300 K obtained
	from DPMD and NPFF simulations. Bottom Row: The first two central
	moments of the iPDFs - variance ($\sigma^2$) and the mean ($\langle
	\mathrm{R} \rangle$) plotted against each other. The color of square
	symbols corresponds to that of the same neighbor in the plots above.
	The white squares show the next nine nearest neighbors. The hydration
	numbers identified from this analysis are indicated above the
	respective square symbols (see SI section S3.3).}
	\label{fig:ipdfs}
\end{figure}

Figure \ref{fig:ipdfs} shows the iPDFs of Cs and Na ions from DPMD and NPFF
simulations. The iPDFs for most O neighbors show Gaussian-like behavior with
large deviations observed near the hydration shell radius (see SI section
S4.1.2). In its bottom row, the peak widths calculated as the
second central moment of the N-th neighbor distance distributions ($\sigma^2$)
are plotted against the average peak position ($\langle \mathrm{R} \rangle$) to
better discern qualitative differences between the hydration shells. The iPDFs
of Cs-O and Na-O  from DPMD show clear qualitative differences, with Cs-ion
having a ``diffuse'' hydration shell while Na-ion possesses a ``structured''
one. The diffuse nature of the Cs-ion hydration shell is evident from the
difficulty in unambiguously assigning a hydration shell radius or hydration
number from the iPDFs. In contrast, the iPDFs of I-O and Cl-O are qualitatively
similar (see SI section S4.1.2). Within NPFF simulations,
iPDFs of Cs-O and Na-O describe structured hydration shells with hydration
number of around six in both cases. Next, we connect these hydration shell
differences between Cs and Na ions to the anomalous diffusion phenomenon by
analyzing the evolution of diffusion coefficients at short times obtained from
velocity autocorrelation functions of various kinds of water molecules (SI
section S3.6).

\begin{figure}[h!]
    \includegraphics[width=1.0\linewidth]{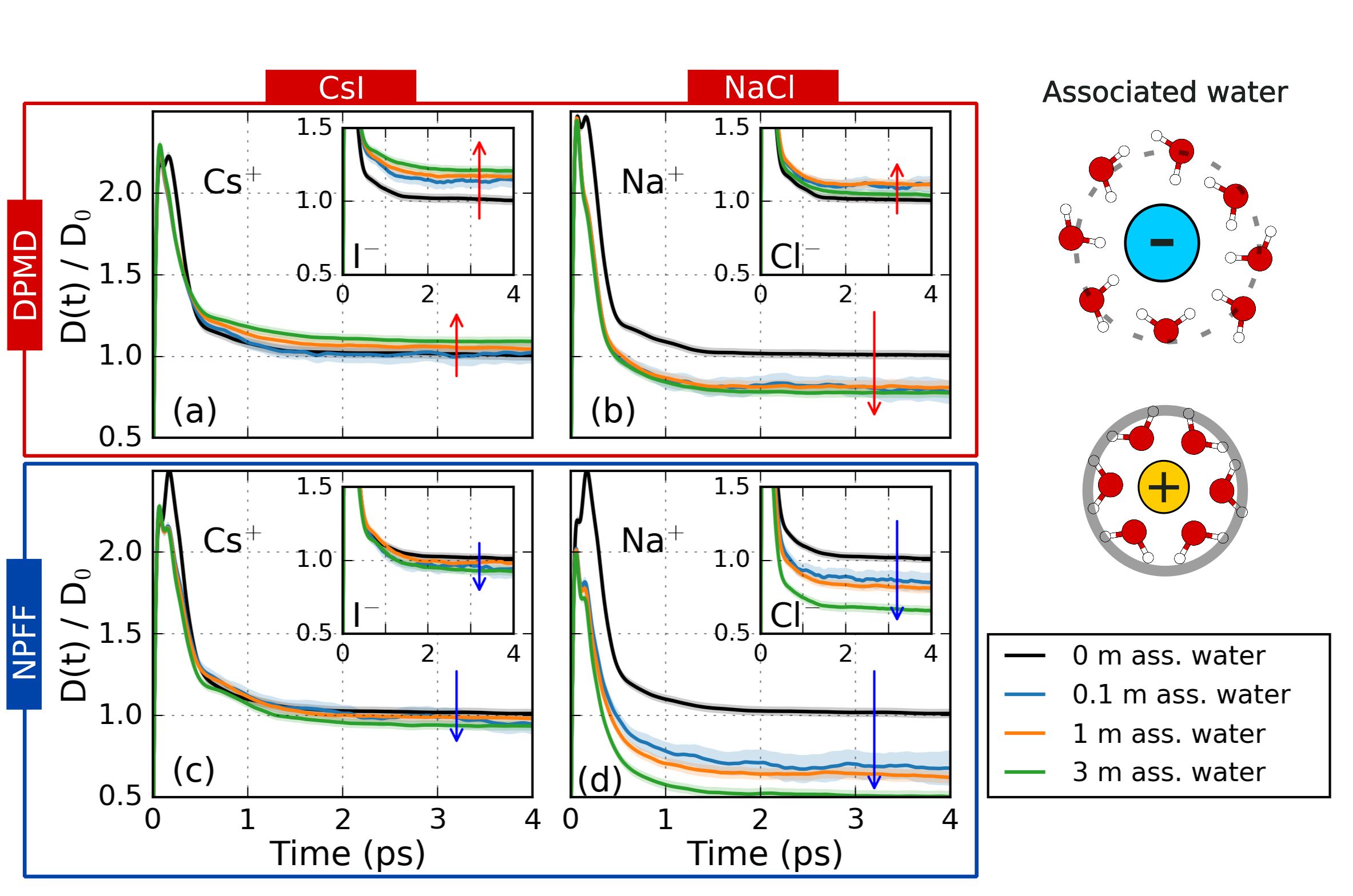}
    \caption{Ratio of running diffusivity (D(t)) of water molecules in aqueous
	solutions to the neat water diffusivity (D$_0$). (a), (b) show the
	diffusivity ratios of cation-associated (at t=0) water molecules in CsI
	and NaCl solutions respectively, obtained from DPMD simulations at 300
	K. (c), (d) show the diffusivity ratios of cation-associated (at t=0)
	water molecules in CsI and NaCl solutions respectively, obtained from
	NPFF simulations at 300 K. Insets show the corresponding plots for
	anion-associated water molecules. The shaded regions around the colored
	lines indicate the 95 \% confidence interval of the corresponding
	diffusivity ratios. The arrows indicate the increment (up-arrow) or
	decrement (down-arrow) of the diffusivity ratios with respect to that
	of neat water. The solvation shell radii were estimated from the first
	minima of the ion-O RDFs, and their values are listed in SI section
	S3.6.} \label{fig:cvacf}
\end{figure}

Figure \ref{fig:cvacf} shows the running diffusivity ratios (D(t)/D$_0$) of
ion-associated water molecules in CsI, NaCl solutions calculated from DPMD and
NPFF simulations. In all cases, the neat water diffusivities at short times (say, at t = 3 ps) 
are close to the corresponding long time diffusivities, as evidenced by their D(t)/D$_0$ values being close to unity. 
Within DPMD, the cation-associated water show varied trends,
with Cs-associated water being faster and Na-associated water being slower than
neat water. On the other hand, both I- and Cl-associated water are faster than
neat water. Hence, the observed differences in the diffusion of water in CsI
and NaCl aqueous solutions can be attributed to the differences in Cs- and
Na-associated water. These results are in accord with the experimental
viscosity B-coefficients of individual ions, with both anions and Cs ion having
negative values while Na ion has a positive value (see SI section
S2.4.1) \cite{collins_methods_2004,marcus_chemrev_2009}. In
contrast, the NPFF simulations show that all ion-associated water molecules are
slower than neat water again highlighting their inability to capture
ion-specific effects.

\begin{figure}[h!]
    \includegraphics[width=0.45\linewidth]{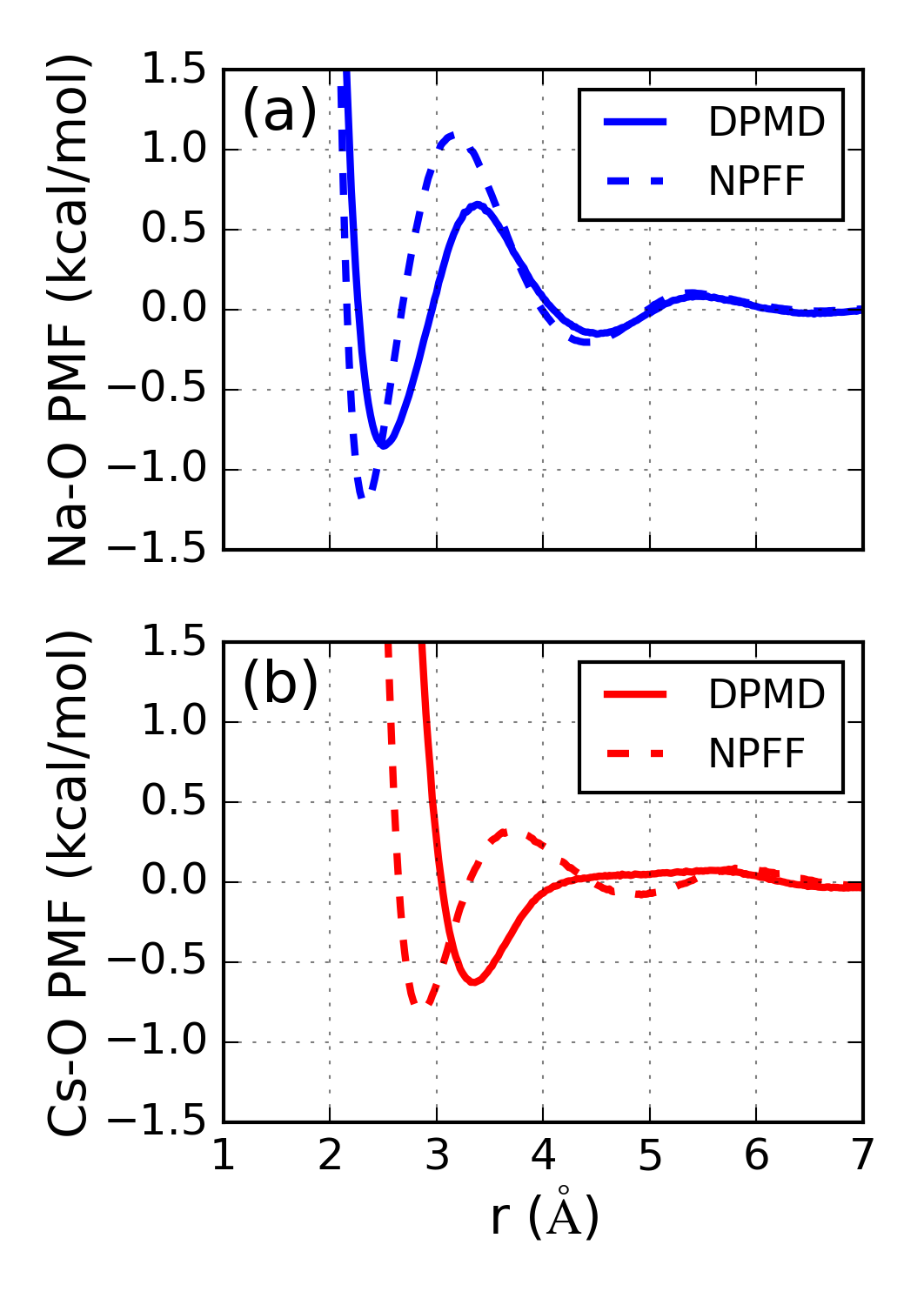}
    \caption{Potential of mean force (PMF) of (a) Na-O pair, (b) Cs-O pair in 3 m aqueous NaCl and CsI solutions respectively at 300 K.}
    \label{fig:pmfs}
\end{figure}

The differences in the behavior of the hydration water can be rationalized
using the potential of mean force (PMF) of the ion-O pairs. The ion-O PMFs
obtained from their respective g(r)s via PMF = -$\mathrm{k_B T}$ln(g(r)), are
generally used to understand the strength of association between them. A common
feature of the cation-O PMFs in aqueous solutions is the presence of a
significant barrier between the potential well (hydration shell region) and the
bulk (solvent region)
\cite{water_exchange_klein_JPCB_2019,jungwirth_JPCL_2023}. Interestingly, the
Cs-O PMF from DPMD does not show a barrier between the hydration shell and the
bulk, thereby facilitating the exchange of hydration and free water molecules
(Figure \ref{fig:pmfs}(b)). In contrast, the Na-O PMF from DPMD shows a clear
barrier (0.7 kcal/mol in 1 m NaCl solution), which differentiates water
molecules belonging to the hydration shell from those in the second shell. This
contrasting behavior between CsI and NaCl solutions is observed across all
concentrations studied in this work (0.1 to 3 m, see Figures
S20-S23). 

On the other hand, both the cation-O PMFs obtained from NPFF simulations are
qualitatively similar, with a barrier separating the hydration shell and the
bulk as seen in Figure \ref{fig:pmfs}. Another difference between DPMD and NPFF
based PMFs is their well depths; those obtained from classical NPFF MD being much
deeper than the corresponding ones from DPMD, indicating that water molecules
in the first hydration shells in the NPFF representation are over-bound. This
phenomenon of over-bound hydration shell by NPFFs is consistent with previous
literature reports
\cite{yethiraj_JPCB_2012,cs_paesani_JPCL_2019,na_k_paesani_JPCB_2022,cl_paesani_JCP_2021,ljfail_bernhardt_JPCL_2022}
and has been primarily attributed to two factors - (i) the lack of explicit
polarizability in their description of ion-water interactions, (ii) the
functional form of the Lennard-Jones term
\cite{yethiraj_JPCB_2012,ljfail_bernhardt_JPCL_2022}.

Finally, we synthesize the DPMD results discussed above to present a coherent
picture of the acceleration-retardation mechanism of water in alkali halide
solutions. Note that the acceleration observed in chaotropic solutions occurs
only at low concentrations and all aqueous solutions eventually show
retardation behavior at high enough concentrations \cite{laage_JACS_2013}.
While the existence of overlapping hydration shells dominate the high
concentration regime \cite{laage_JACS_2013}, ion-specific effects dominate at
low concentrations
\cite{collins_methods_2004,marcus_chemrev_2009,laage_JACS_2013}. The following
discussion mainly pertains to the latter regime.

First, the central effect of ions on the dynamics of water molecules is
restricted to the first few solvation shells and water molecules present
outside these shells behave like bulk-water (see SI section
S4.3.1) \cite{bakker_science_2003,laage_JACS_2013}. Hence,
the main contribution to the acceleration-retardation mechanism comes from the
solvation shell water which is markedly different for NaCl and CsI solutions
(Figure \ref{fig:cvacf}). Second, the contribution of cations and anions to the
acceleration-retardation mechanism is quite different in CsI and NaCl
solutions. While both ions contribute to the acceleration mechanism in CsI
solutions, the retardation observed in NaCl solutions is the result of a
delicate (temperature-dependent) competition between the retardation caused by
Na-associated water and the acceleration caused by Cl-associated water. This
competition between ions is particularly germane in light of a recent
observation that supercooled NaCl solutions are chaotropic
\cite{nacl_chao_yethiraj_JPCB_2008,nacl_chao_corti_JCP_2011,nacl_chao_price_JPCB_2014}.
This argument is also supported by the sign of viscosity B-coefficients, with
Cs/I having large negative values, Na having a large positive value and Cl
having a near-zero value (see SI section S2.4.1)
\cite{collins_methods_2004,marcus_chemrev_2009}. Finally, the dynamics of
hydration water can be connected to the hydration shell structure, with Cs/I
ions having a qualitatively more disordered one than that of Na/Cl ions (See SI
section S2.3)\cite{parrinello_PNAS_2014}.

In summary, we trained MLPs for aqueous NaCl and CsI
solutions using revPBE-D3 based DFT data on liquid phase configurations. These
MLPs are shown to be good models of the physical salt solutions in reproducing
various thermodynamic, structural and transport property benchmarks (Figure
\ref{fig:benchmark}, Table \ref{tab:benchmark}, and SI section
S2). Crucially, DPMD simulations based on these MLPs can
capture the ``anomalous water diffusion'' phenomenon which has been an
outstanding challenge to most NPFFs. The microscopic
origins of this phenomenon in CsI solutions is shown to lie in the ``diffuse''
hydration shell of Cs ions, which enables a faster movement of water molecules;
in contrast, the ``structured'' hydration shell of Na ions in NaCl solutions is
responsible for the observed slowdown. The hydration shells described by NPFF
simulations are (i) over-bound and (ii) more compact and hence do not
capture this phenomenon. The quantitative reproduction of transport properties
such as diffusivity and shear viscosity is non-trivial and to do so at ab
initio accuracy has been possible only because of MLPs' ability to extend
ab initio accuracy to large-scale MD simulations.  

\section{Computational Methods}
The Machine Learned force fields were constructed using the Deep Potential
framework which uses an end-to-end neural network representation
\cite{deepmdkit_CPC_2018,deepmdse_NIPS_2018,deepmdkit_arxiv_2023}.
Specifically, we use the short-range version with 6 $\mathrm{\AA}$ cutoff for
the local atomic representations. Further details about the MLP architecture
and the training procedure are outlined in SI section S1.
The quantum mechanical data, on which the MLPs were trained, was obtained from
periodic DFT calculations of condensed phase snapshots. These single point
energy calculations were carried out using the revised PBE exchange correlation
functional \cite{PBE_1996,revPBE_1998} including the Grimme’s dispersion
correction \cite{grimme_D3_2010} (revPBE-D3). We found that the revPBE-D3
functional has a good balance between accuracy and computational cost when
applied to water and aqueous solution systems \cite{angelos_arxiv_2022}.
Further details about the DFT calculations, the choice of the functional and
benchmarks are presented in SI sections S1 and S2.
The condensed phase structures used to train the MLPs
were sampled using an active learning procedure like that in DP-GEN
\cite{dpgen_CPC_2020}, details of which are provided in SI section
S1.3. The training of the MLPs was carried out using the
DeepMD-kit package \cite{deepmdkit_CPC_2018}. All the quantum calculations were
performed using CP2K \cite{cp2k_JCP_2020}. 

MD simulations using MLPs were run using LAMMPS \cite{lammps_CPC_2022} patched
with DeepMD-kit. Isotropic isobaric-isothermal (NPT) DPMD simulations were
carried out for 2 ns to obtain the equilibrium densities of the electrolyte
solutions. Fully flexible barostat was used to obtain the equilibrium density
of ice-\textrm{I}h. In both cases, Nos\'e-Hoover thermostat and barostat (as
implemented in LAMMPS), with time constants of 0.5 and 1.0 ps respectively,
were used to control the temperature and pressure respectively
\cite{barostat_shinoda_PRB_2004}. Subsequently, the DPMD production runs were
carried out in the canonical ensemble (NVT) with Nos\'e-Hoover thermostat. A
timestep of 0.5 fs was used to integrate the equation of motion for all DPMD
simulations. The recently developed Madrid-2019 force field was used as a
representative of classical NPFF for aqueous
solutions
\cite{vega_tip4p2005_JCP_2005,vega_madrid2019_JCP_2019,vega_madrid2019_JCP_2022}.
All the NPFF simulations were carried out using the GROMACS package
\cite{gromacs_softwarex_2015}.

Estimating transport properties from equilibrium MD simulations is non-trivial,
generally requiring long trajectories with multiple independent runs
\cite{livejournal}. This situation is further complicated by their sensitive
dependence on simulation parameters like - simulation box size, thermostat
parameters, the choice of estimators, long range interactions, etc. In this
work, we carefully test the robustness of the estimates wherever possible (see
SI section S3), and some important rationale regarding the
choice of the simulation parameters used in the estimation process are
described below. (1) \textit{Box size dependence:} The self-diffusion
coefficients estimated from simulations with periodic boundary conditions have
a significant box size dependence (D $\sim 1/L$) \cite{yeh_hummer_JPCB_2004}.
However, the ratios of diffusion coefficients D/D$_0$ do not show a box size
dependence as demonstrated in SI section S3.4.3 and consistent
with previous observations \cite{yethiraj_JPCB_2012,panagiotopolous_JPCB_2023}.
(2) \textit{Thermostat parameters:} Though the application of some thermostats
can alter the estimates of diffusivity and shear viscosity, Nos\'e-Hoover
thermostat (used in this work) was shown to have minimal impact on the same
\cite{shirts_thermostat_JCTC_2013,klein_thermostat_JPCB_2022}. (3)
\textit{Estimators:} While there exist many estimators for the estimation of
transport properties
\cite{mandadapu_JCP_2012,livejournal,octp_JCIM_2019,hummer_optimal_est_JCP_2020,baroni_NPJComp_2022},
it is not entirely clear if there are systematic differences between them.
Here, we use the time decomposition method in combination with bootstrap
procedure, which is generally regarded as a good practice, to estimate the
transport properties \cite{livejournal,bala_JCTC_2021}. (4) \textit{Long range
interactions:} Though the explicit inclusion of long range interactions into
the DP framework is necessary to capture the liquid-vapor properties, bulk
properties like density, diffusivity, and viscosity were shown to be estimated
accurately from short range DP models
\cite{panagiotopolous_SR_JCP_2021,yue_thesis_2021}.

\begin{acknowledgement}
The support and the resources provided by ``PARAM Yukti Facility'' under the
National Supercomputing Mission (NSM), Government of India, at the Jawaharlal
Nehru Centre For Advanced Scientific Research are gratefully acknowledged by
Nikhil Avula (N.A.) and Sundaram Balasubramanian (S.B.) This work is partly
funded by the NSM project entitled ``Molecular Materials and Complex Fluids of
Societal Relevance: HPC and AI to the Rescue'' (Grant no.
DST/NSM/R$\&$D\_HPC\_Applications/2021/05). N.A. thanks Dr. Nimish Dwarkanath
for his assistance in running some DPMD simulations. 
\end{acknowledgement}

\begin{suppinfo}
Supporting Information file can be freely accessed from: https://doi.org/10.5281/zenodo.8192558
\end{suppinfo}

\bibliography{ref}

\end{document}